\documentclass[preprint,11pt,number]{elsarticle}

\usepackage[a4paper,margin=2cm]{geometry}

\usepackage[T1]{fontenc}
\usepackage[utf8]{inputenc}

\usepackage{amsmath,amssymb,amsfonts,mathtools}
\usepackage{amsthm}
\allowdisplaybreaks[4]
\numberwithin{equation}{section}

\usepackage{graphicx}
\graphicspath{{Figures/}{Figures/}}

\usepackage{booktabs}
\usepackage{threeparttable}
\usepackage{subfig}

\usepackage{algorithm}
\usepackage{algpseudocode}

\usepackage[dvipsnames]{xcolor}
\usepackage{url}
\usepackage[
    colorlinks=true,
    linkcolor=blue,
    citecolor=blue,
    urlcolor=blue
]{hyperref}
\usepackage[nameinlink,noabbrev]{cleveref}

\newcommand{\rev}[1]{#1}

\theoremstyle{plain}

\theoremstyle{definition}

\theoremstyle{remark}

\begin{document}

\begin{frontmatter}

\title{Fairness, Travel, and Market Potential:
An Optimization Framework for NBA Expansion}

\author[manchester]{Ali Hassanzadeh\corref{cor1}}
\ead{ali.h@manchester.ac.uk}

\author[skema]{Morteza Davari}

\author[ghent,flanders]{Dries Goossens}

\cortext[cor1]{Corresponding author}

\address[manchester]{Centre for AI and Decision Sciences,
Alliance Manchester Business School,
University of Manchester, UK}

\address[skema]{SKEMA Business School,
Universit\'e C\^ote d'Azur, France}

\address[ghent]{Department of Business Informatics
and Operations Management,
Ghent University, Belgium}

\address[flanders]{FlandersMake@UGent -- core lab CVAMO,
Belgium}

\begin{abstract}
The potential expansion of the National Basketball Association (NBA) to 32 teams raises the question of how to restructure its conferences and divisions while balancing travel efficiency, fairness in team travel burdens, and commercial exposure.
This study provides a quantitative framework for expansion planning.
We develop two optimization models: one minimizing total travel distance and another following the Kalai--Smorodinsky (KS) proportional-improvement principle. The KS model maximizes the smallest normalized travel improvement between a team's fixed reference and individual ideal. Both models incorporate an opponent-exposure constraint to account for commercial considerations, and take into account road trips when measuring travel distance.
Using data from all 30 current franchises and six candidate cities (Seattle, Las Vegas, Montreal, Vancouver, Tampa, and Mexico City), we evaluate 15 pairwise expansion scenarios under alternative season lengths and divisional formats.
In the 30-team comparison, KS strikes a good balance between overall travel efficiency and fairness across teams, giving greater attention to individual travel burdens while keeping total travel close to the distance-minimizing solution. The expansion results show that travel requirements depend jointly on the candidate city pair and the competition format, highlighting the need to assess expansion and realignment together.
Our study offers a flexible decision support framework for league executives, policymakers, and sports economists. It makes the trade-offs between total travel, individual gains, and opponent exposure explicit, providing evidence-based guidance for strategic league design.
\end{abstract}


\end{frontmatter}
\section{Introduction}\label{sec:intro}

Since its formal establishment in 1949, the National Basketball Association (NBA) has undergone significant transformation. The early years were marked by frequent team relocations and contractions, but from the late 1960s onward the league steadily expanded. The most recent addition came in 2004, when the \emph{Charlotte Bobcats} (now the \emph{Charlotte Hornets}) joined, bringing the total to 30 teams. Since then, the structure has remained unchanged, even as the sport has become more global and more profitable.

In recent years, NBA expansion has re-emerged as a central topic. Reports have highlighted cities such as Seattle, Las Vegas, Vancouver, and Mexico City as strong candidates, given their fan bases, infrastructure, and growing interest in basketball \citep{guardian-expansion}. In 2024, Commissioner Adam Silver confirmed expansion would be considered once the new media rights deal was complete \citep{nyt-expansion}, and franchise valuations have since surged, as seen in the \$6.1 billion sale of the Boston Celtics in early 2025 \citep{forbes-sale}, with expansion teams expected to cost between \$4 and \$7 billion \citep{forbes-sale, ministryosport-expansion}. Together, these developments underscore that expansion is no longer a question of \emph{if}, but of \emph{how}.

Expanding the NBA to 32 teams raises a key question: how should conferences and divisions be realigned? Since 2004, the league has operated with two conferences, each divided into three divisions of five teams. Adding two new franchises forces a redesign that will affect logistics, fairness, costs, player performance, and revenue.

One of the biggest challenges is travel. Teams in remote markets often log thousands more miles, and studies consistently show that extensive travel, especially across time zones, reduces performance \citep{entine2008,ashman2010,roy2017,huyghe2018,mchill2020,charest2021}. Reducing these inequities is therefore central to any realignment plan.
%
%
Sport-related travel also matters for sustainability, through air travel drives emissions. Reducing total league-wide travel therefore lowers both player fatigue and the NBA’s carbon footprint, an increasingly important goal for global sports organizations \citep{Menge2025}.
\rev{Our focus on NBA expansion is long-term. Accordingly, rather than modeling the details of any particular season, we focus on the league’s conference and division structure. We account for the possibility of combining away games into road trips, but do not construct a dated schedule.
}
%

%
\rev{Beyond travel, divisional design changes how often franchises face particular opponents. Commercial appeal depends on more than city population, including a franchise's brand and wider audience. We therefore use reported incumbent franchise valuations to construct an opponent-exposure measure and require incumbent teams to retain a specified proportion of their baseline exposure. This is a commercial safeguard, not a franchise revenue forecast or a revenue-sharing model.
}


%
\rev{
}

Finally, we highlight our contributions. Operations research has long supported policy and design decisions in professional sports \citep{prince2012,smith2008,yi2020}. Inspired by this line of work, we propose two integer programming formulations that model divisional realignment following the addition of two expansion teams. The first model minimizes total league-wide travel distance\rev{, as a benchmark. A reduction in league-wide travel may however benefit some teams while increasing the travel of others. We therefore introduce a Kalai--Smorodinsky (KS) model maximizing the worst normalized improvement between fixed travel references and individual ideals.
Both models account for travel distance based on road trips, rather than considering only the distances between teams within and across divisions and conferences.
} The result is a decision support tool that evaluates expansion scenarios under alternative city choices, divisional formats, and season lengths, providing actionable insights for league executives.



\section{Competition Design}
\label{sec:comp-design}

Competition (or tournament) design refers to the rules that determine how a tournament, a series of games between a number of competitors, is organized. These rules determine the set of contestants, the format, the schedule, the ranking, and the prize allocation \citep{devriesere2024}. \citet{devriesere2024} discuss the impact of the competition design on efficacy (do strong teams rank high?), fairness (are teams treated equally?), attractiveness (how much excitement does the competition generate?), and strategy-proofness (do teams have an incentive to exert full effort?). Moreover, numerous studies have examined how competition design and league policies influence attendance and revenue. In the case of the NBA, research has explored, e.g.\ the effects of the salary cap \citep{Chatzistamoulou2022}, the playoff structure \citep{LongleyLacey2012}, the COVID-19 ``bubble'' setup \citep{Hindman2021}, changes in collective bargaining \citep{Bognar2024}, and the draft system \citep{Lenten2016}.

This paper focuses on the set of contestants and the format, i.e.\ the set of matches to be played. The two most common formats are the round robin tournament and the knockout tournament. In the former, each team faces each other team the same number of times. In the latter, teams face only a subset of the other teams, and are immediately eliminated if they lose. For the NBA, neither format is suitable: a round-robin tournament would require a lot of travel time between teams on opposite sides of the country. A knockout tournament, on the other hand, would offer too few games and eliminate many teams very quickly. Hence, the NBA opted for a hybrid tournament. In the regular season, each team faces each other team two to four times, while the subsequent playoffs (and play-in) are organized as a knockout tournament among the best teams of the regular season.

The NBA regular season spans approximately 180 days, running from October to April each year. During this period, each team plays 82 games, averaging one game every two days. Although the number of teams in the league has fluctuated over time, the 82-game requirement per team has remained unchanged since the 1967-68 season, with four exceptions due to extraordinary circumstances such as player strikes and the COVID-19 pandemic. The league structure determines the frequency of matchups between teams. Each team plays four games against each of the four other teams within its division ($4 \times 4=16$ games), four games against six teams from other divisions within the same conference ($4 \times 6=24$ games), and three games against the remaining four teams in its conference ($3 \times 4=12$ games). Additionally, each team competes in two games against each of the 15 teams from the opposing conference ($2 \times 15=30$ games). A five-year rotational system dictates which out-of-division conference opponents are played only three times per season.
\rev{Each out-of-division conference opponent is played four times in three seasons and three times in two seasons, giving $4+4+4+3+3=18$ games over the five-season rotation; the order depends on the opponent's position in the rotation.}
Over the course of five seasons, each team will have played 20 games against every in-division opponent, 18 games against each out-of-division conference opponent, and 10 games against each team from the opposing conference.

\begin{figure}[ht]
    \centering
    \vspace{-10pt}
        \includegraphics[width=1\textwidth]
    {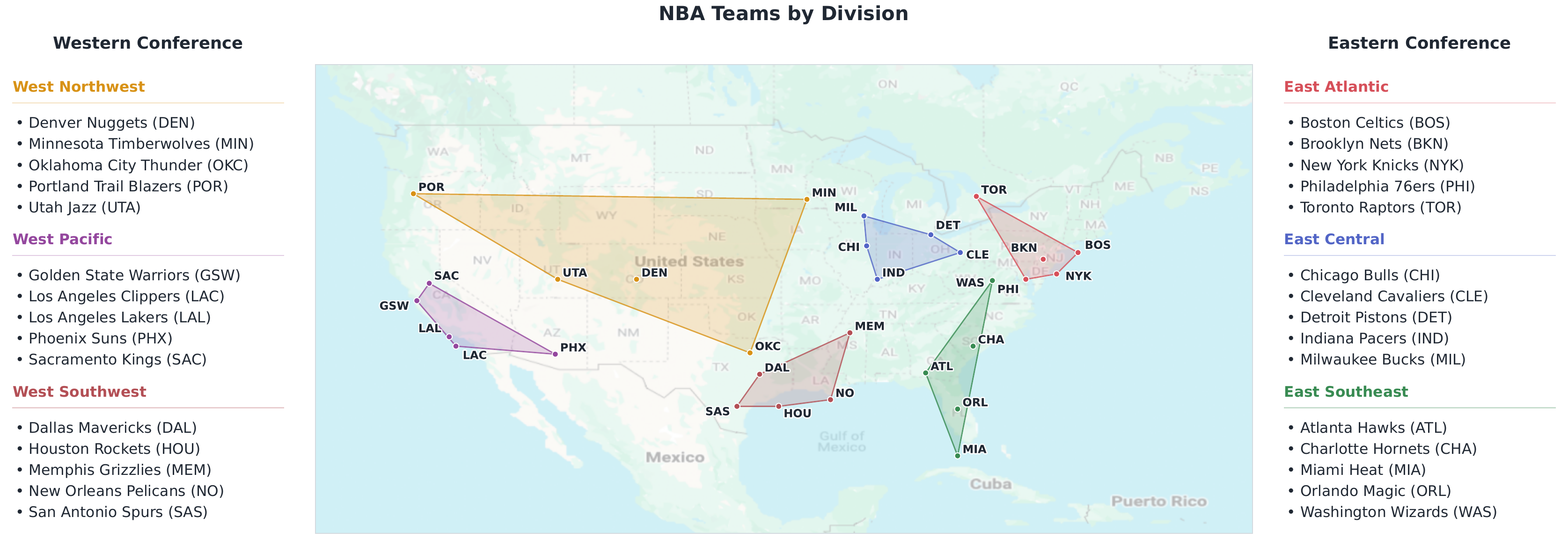}
    \caption{Assignment of teams to divisions and conferences in the current format of NBA.}
    \label{fig:nba-teams}
\end{figure}

A crucial decision is the grouping of teams into divisions and conferences, as it affects competitive balance, the attractiveness of the competition, as well as the travel distance teams must cover. This problem is known as the sport teams (re)alignment \citep{mitchell2003} or grouping \citep{toffolo2019} problem. \citet{ji2005} develop a method to find a team realignment that minimizes the total intradivisional travel, and find that the NBA's solution (in 2004) is very close to that optimum (0.089\% worse).
%
\rev{Later, \citet{macdonald2014} develop a realignment method using a matchup-weighted city-to-city travel measure. Under this measure, they report that the NBA alignment examined in their study entails approximately 100 miles more total league travel than their best alternative alignment.}
A complete list of teams, along with their current assignment to divisions and conferences, is presented in Figure \ref{fig:nba-teams}.

It is important to realize that a team's travel over a season is determined not only by i) the locations of the teams assigned to its division, its conference, and the other conference, but also by ii) the number of away games against those teams. A third major factor is the schedule, because teams can make trips connecting consecutive away games (instead of returning home after each away game). While the tournament format decides which games are to be played, the schedule decides on the timing (and order) of the games. Scheduling a league is a far from trivial task, where typically many, often conflicting, wishes and requirements of various stakeholders have to be taken into account \citep{vanbulck2020}. Several studies have explored the role of operations research in scheduling professional basketball competitions, such as those in Argentina \citep{duran2019}, the Czech Republic \citep{froncek2001}, Germany \citep{westphal2014}, and New Zealand \citep{wright2006}.

%
\rev{Travel-oriented scheduling has also been studied through the traveling tournament problem \citep{easton2001, vanbulck2026TTPoverview}, which abstracts from many practical scheduling requirements to focus on travel minimization. In the NBA context, \citet{bean1980} developed scheduling methods to reduce airline miles during the 1978--79 and 1979--80 seasons, when the league consisted of 22 teams.
Our study concerns conference and division realignment, which determines opponent frequencies and the opportunities for combining away games into road trips. We model these travel consequences over a multi-season rotation, while leaving the construction of a dated schedule and its operational requirements outside the scope of the analysis.}

%
\rev{
Competition expansion can give rise to concerns over how the resulting benefits are allocated. \citet{krumer2023allocation} study the allocation of additional FIFA World Cup places among confederations using claims-based rules, while \citet{csato2025allocation} propose allocations based on confederation rankings. \citet{osicka2023fair} use cooperative game theory to account for fairness in tournament travel. These studies motivate explicit treatment of stakeholder interests and fairness. 

Individual ideal travel distances can be interpreted as aspirations that generally cannot be achieved simultaneously. This links our setting to claims problems, where a scarce resource or endowment is smaller than the sum of the individual claims placed on it, requiring a fair rule to divide the available amount among the claimants \citep{thomson2003axiomatic, Thomson2019}. Unlike a classical claims problem, however, the available travel savings do not constitute a fixed estate: changing the alignment affects both total travel and the distribution of achievable savings across teams. 
}

\section{Methodology}
\label{sec:fairness-in-leagues}

In this section, we present the methodological framework used to evaluate the impact of NBA expansion and to determine optimal realignments of teams into divisions and conferences.
Two key dimensions are considered throughout. First, travel distance plays a central role, as it constitutes one of the most important operational costs in the NBA. Section~\ref{sec:measure-travel-distance} explains how travel distance is measured, accounting for both road trips and matchup rotation across multiple seasons. Excessive travel not only increases financial and environmental costs but also contributes to player fatigue and reduced performance.
Second, we account for the revenue implications of realignment with respect to opponent exposure (see Section~\ref{sec:measure-rev-impact}).
%
\rev{
Section~\ref{sec:min-total-distance} presents a model that minimizes total travel distance and serves as an efficiency benchmark. Section~\ref{sec:ks-model} then develops the KS proportional-improvement formulation.
}

\subsection{Measuring travel distance}\label{sec:measure-travel-distance}

The distance traveled by NBA teams in a given season is determined by the league schedule. Because schedules vary from year to year, the actual distances covered by teams also fluctuate across seasons. As discussed in Section~\ref{sec:comp-design}, however, constructing schedules is computationally complex, and realistic inputs are not available. While these season-specific figures are operationally important, our focus is strategic: we abstract from annual schedule variation and examine how the underlying divisional and conference structure shapes travel distances over the long run. To this end, we use a travel-distance measure that can be computed efficiently and serves as a surrogate for the travel efficiency of a league structure.

\rev{As far as we are aware, the literature on the NBA (re)alignment problem uses a distance measure that assumes that each team returns to its home venue after each away game \cite{ji2005,macdonald2014}, which we refer to as the direct-distance proxy. Given an alignment of teams to divisions and conferences, the direct-distance proxy $r_i^{\mathrm{dir}}$ is defined for each team $i$ as

\begin{equation}
r_i^{\mathrm{dir}} = \sum_{j\in D_i\setminus \{i\}} d_{ij}*n_1 + \sum_{j\in C_i\setminus D_i} d_{ij}*n_2 + \sum_{j\in T\setminus C_i}d_{ij}*n_3,
\end{equation}
with $T$ the set of all the teams in the NBA, $D_i$ ($C_i$) the set of teams in the same division (conference) as team $i$, and where $d_{ij}$ is the distance between teams $i$ and $j$, and $n_1, n_2, n_3$ denote the (long-term average) annual number of games played against divisional, intra-conference, and inter-conference opponents, respectively. Currently, $n_1=4, n_2=3.6$, and $n_3=2$ (see Section~\ref{sec:comp-design}); the assignment of teams to divisions and conferences is given in Figure~\ref{fig:nba-teams}.

In this paper, we propose a refined distance measure that accounts for road trips. Road trips are common in NBA schedules and can substantially reduce travel by allowing teams to play consecutive away games before returning home, rather than making a separate round trip for each game.
A road trip $p=(i,j_1,\ldots,j_{\ell_p},i)$ starts and ends at team $i$'s home location and has distance
\begin{equation}
c_{ip}
=
d_{ij_1}
+\sum_{k=1}^{\ell_p-1}d_{j_kj_{k+1}}
+d_{j_{\ell_p}i}.
\label{eq:trip-distance}
\end{equation}

A road-trip-based distance measure, however, requires an integer matchup frequency for each opponent, which is not satisfied since $n_2=3.6$. Therefore, we determine a rotation horizon $H$ such that $a_k=Hn_k/2$ is integer for each category $k \in \{1,2,3\}$.
The horizon $H$ counts seasons in the rotation. For the current 30-team format, $H=5$, giving integer away-visit totals of $(a_1,a_2,a_3)=(10,9,5)$ per opponent. Dividing travel over the rotation by $H$ gives an average annual distance.
The required number of away visits of team $i$ to opponent $j$ over the rotation horizon is then

\begin{equation}
\left\{
\begin{aligned}
v_{ij} &= a_1 \textrm{ if } j\in D_i\setminus \{i\},  \\
v_{ij} &= a_2 \textrm{ if } j\in C_i\setminus D_i,  \\
v_{ij} &= a_3 \textrm{ if } j\in T\setminus C_i.  \\
\end{aligned}
\right.
\label{eq:horizon-visits}
\end{equation}

For the current NBA structure, $H=5$, and $a_1$, $a_2$, and $a_3$ equal 20, 18, and 10, respectively.


For team $i$, let $P_i$ be the set of candidate road trips, 
and let $\delta_{jp}$ count visits to opponent $j$ in trip $p$. We now define a variable  $z_{ip}$ which represents the number of times that trip $p$ is used by team $i$ over the rotation. The road trip distance $r_i$ for each team $i$ is then obtained through equation (\ref{eq:distance-def}) in the following model:

\begin{align}
\text{min} \quad &\sum_{i \in T} r_i&&\\
\text{s.t.} \quad &r_i = \frac{1}{H}\sum_{p\in P_i}c_{ip}z_{ip} &&i\in T,\label{eq:distance-def}\\
&\sum_{p\in P_i}\delta_{jp}z_{ip}=v_{ij}, &&i\ne j,\label{eq:trip-cover0}\\
&z_{ip}\in\mathbb Z_{\ge0}, &&i\in T,\ p\in P_i.
\end{align}


Trip length counts the number of away games before returning home. For trips of one to three games, we generate all opponent combinations and retain the shortest visiting order for each. Longer trips are sampled from historical itineraries. For expansion scenarios, we also generate trips by inserting an expansion opponent into an existing trip, increasing its length by one game. The generated pool is then pruned to keep the optimization manageable.
Bounds on trip counts limit the concentration of away visits into a small number of long tours. Optimization guarantees depend on the retained candidate pool. Note that the model does not guarantee season-level feasibility. Furthermore, synchronization of calendar dates, arena availability, rest requirements, and broadcast scheduling remains outside its scope.


In order to validate our road-trip measure, we compare it with the reconstructed actual travel and the direct-distance proxy. The reconstructed actual travel follows the sequence of historical game venues. The comparison uses the same 30 teams over four seasons, from 2015-16 to 2018-19, with all distances expressed in miles.

For the correlation analysis, each measure is first averaged within team over the four seasons. Figure~\ref{fig:travel-correlations} reports Pearson correlation $r$, which measures linear association, and Spearman correlation $\rho$, which measures agreement in team rankings. The road-trip measure has stronger association with reconstructed travel ($r=0.90$, $\rho=0.90$) than the direct-distance proxy ($r=0.67$, $\rho=0.63$). These results support accounting for road-trip structure when comparing travel burdens across teams.

\begin{figure}[!htbp]
\centering
\includegraphics[width=0.6\textwidth]
{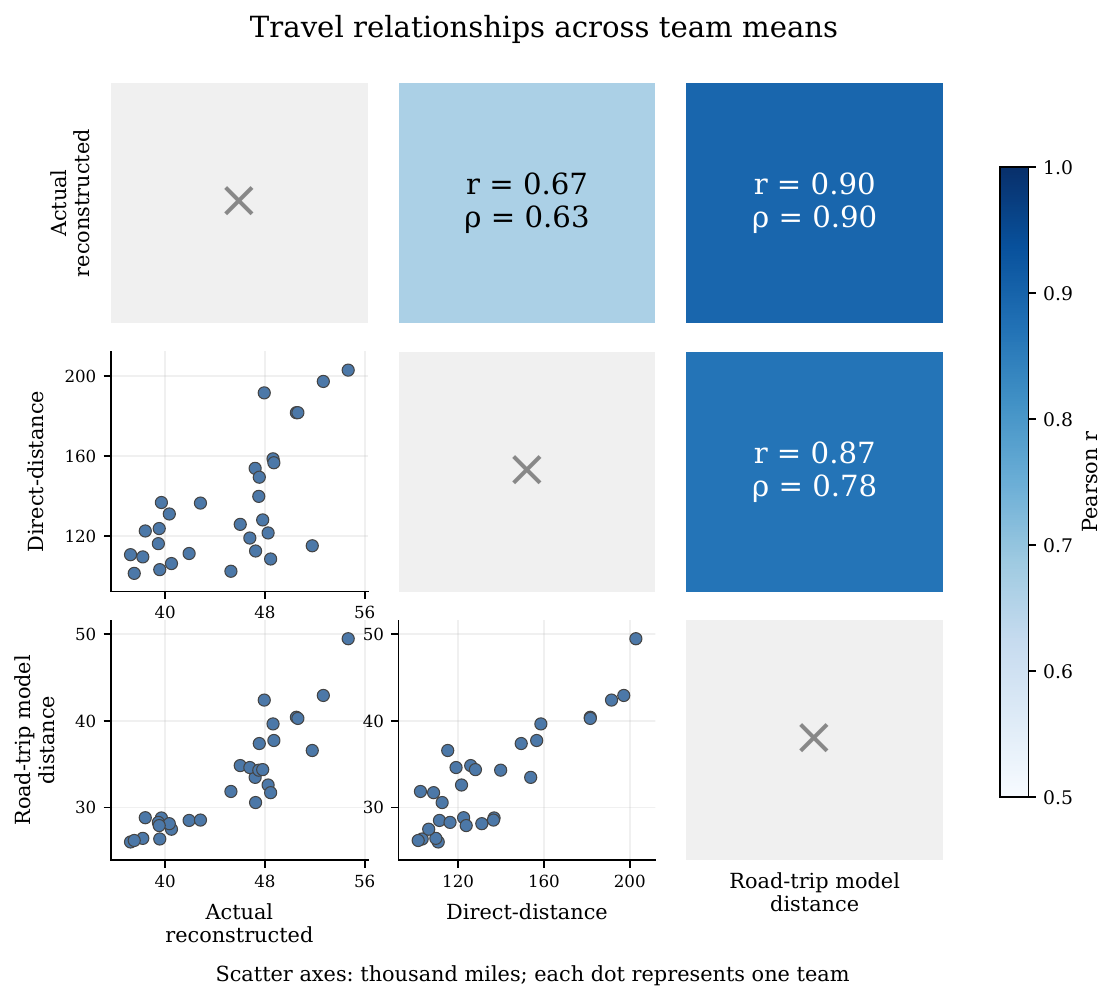}
\caption{Relationships between reconstructed actual travel, the direct-distance proxy, and road-trip model travel across 30 team means. Lower-triangle panels show scatterplots, with each dot representing one team. Upper-triangle panels report Pearson $r$ and Spearman $\rho$; colour represents Pearson $r$.}
\label{fig:travel-correlations}
\end{figure}

Figure~\ref{fig:travel-validation} compares actual seasonal variation with the direct-distance and road-trip model means. The direct-distance model, which ignores the possibility of road trips, substantially overestimates actual travel. In contrast, the road-trip model systematically underestimates reconstructed actual travel. This is because the road-trip measure assumes that away visits can be optimally grouped within the candidate pool, whereas realized schedules are subject to additional operational requirements.
Nevertheless, the strong correlations among historical team means and the road-trip model results indicate that this model broadly preserves the relative travel burdens observed across teams and support its use for comparative analysis. We therefore adopt the road-trip model and use $r_i$ to compare travel burdens across various divisional and conference structures, rather than as an exact prediction of realized seasonal mileage. 



\begin{figure}[!htbp]
\centering
\includegraphics[width=\textwidth]
{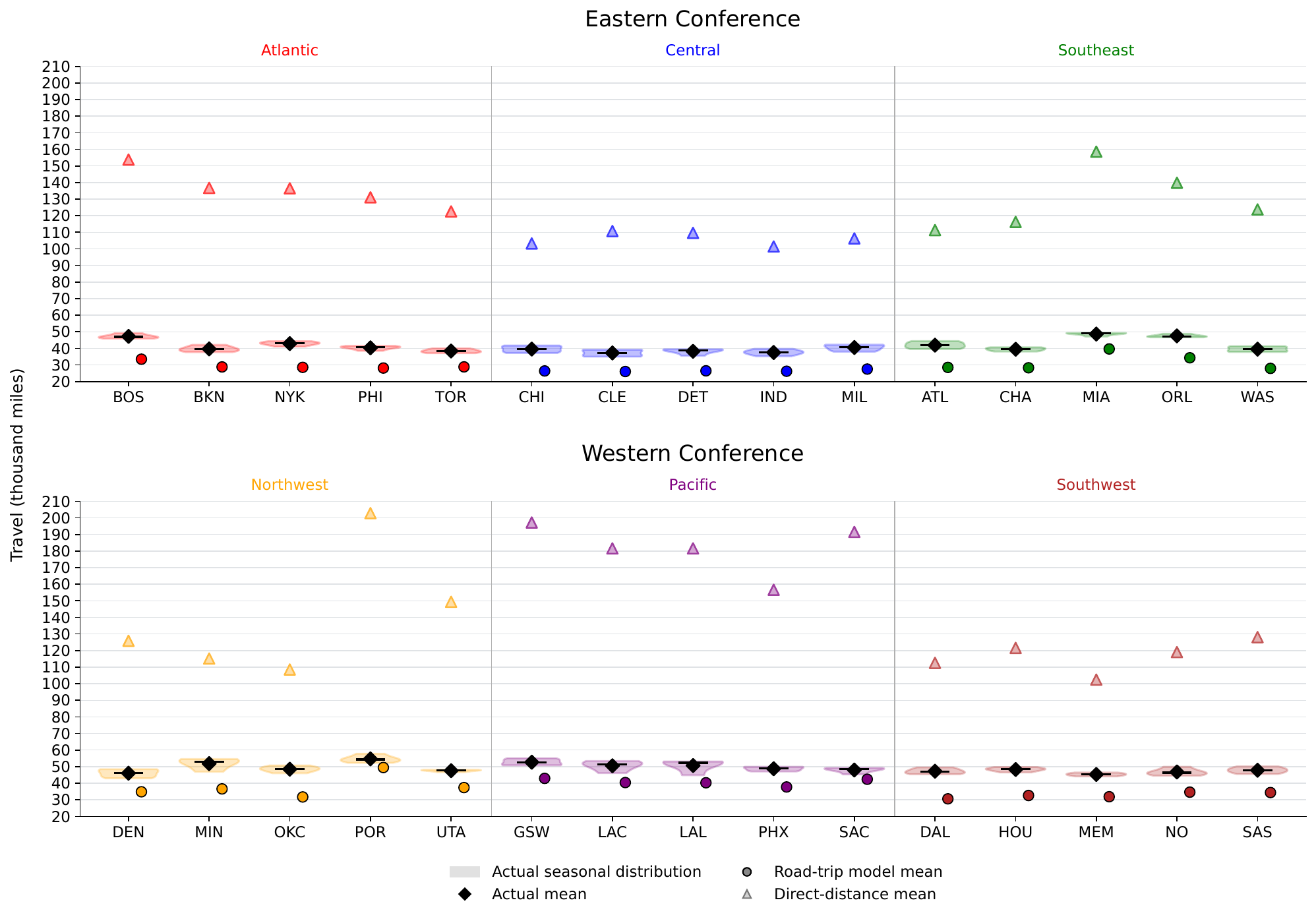}
\caption{Reconstructed actual travel and direct-distance and road-trip model estimates. Violins show smoothed distributions of the actual seasonal distances for each team; horizontal lines mark medians and black diamonds mark means. Coloured triangles (circles) show direct distance (road-trip model) means over the same seasons. Colours identify divisions, and both conference panels use the same distance scale.}
\label{fig:travel-validation}
\end{figure}

}

%
\rev{\subsection{Measuring commercial opponent exposure}\label{sec:measure-rev-impact}}

One of the central questions in evaluating NBA expansion is how the introduction and positioning of new teams affects the \rev{
incumbent teams in terms of commercial opponent exposure. We derive an opponent-exposure measure from published franchise valuations and impose a lower bound on the share of their initial exposure that incumbent teams must preserve.

For incumbent team $j$, $m_j$ is its Forbes 2025 franchise valuation in USD billions \citep{forbes2025values}. Each expansion candidate is assigned the incumbent median, USD 4.75 billion. This common imputation avoids claiming observable franchise values for teams that do not yet exist. The six candidate cities are therefore distinguished by travel geography in the base case, rather than by estimated candidate-specific franchise values.

For an assignment of teams to divisions ($x$) and conferences ($y$), we define team $i$'s valuation-weighted annual opponent exposure as follows:
\begin{equation}
M_i(x,y)=\sum_{j\in T\setminus\{i\}}m_j(2v_{ij}/H).
\label{eq:exposure-def}
\end{equation}
where $2v_{ij}/H$ corresponds to the average annual matchup frequency between team $i$ and $j$, which depends on the composition of division and conferences.

An incumbent opponent's weight is its reported franchise valuation, which reflects commercial characteristics. Franchise valuation remains an imperfect indicator of a visiting team's appeal: it also reflects assets, expected growth, and scarcity.

The valuation-weighted opponent exposure isolates the opponent component that the alignment changes (i.e.\ how frequently each team faces each of its opponents), rather than predicting franchise revenue or estimating the financial return from expansion. Actual revenues also depend on own-market characteristics and revenue-sharing arrangements, neither of which estimated by this measure. Although the expansion may, in the long run, impact the teams' own valuation, we do not take this into account here. Broadcasting revenue allocation \citep{bergantinos2020sharing} is a separate issue that falls outside the scope of this paper; franchise valuations do not determine payments under the NBA's revenue-sharing arrangements.
Finally, teams within the same division need not have identical exposure scores: a less attractive team, for instance, may benefit far more from playing against a high-profile team than the reverse. Moreover, similar exposure scores do not necessarily translate into similar revenues.

Let $T^I$ be the incumbent teams and $\bar M_i$ their exposure under the pre-expansion 30-team, 82-game structure, evaluated using the same valuations. For a $G$-game season, we use the scaled benchmark $\bar M_i=(G/82)\bar M_i^{82}$ and impose
\begin{equation}
M_i(x,y)\ge\gamma\bar M_i,\qquad \forall i\in T^I,
\label{eq:market-constraint}
\end{equation}
with $\gamma=0.8$. This ensures that incumbent teams retain at least 80\% of their baseline opponent exposure after adjusting for season length. Expansion teams have no pre-expansion benchmark and are not subject to this constraint. We treat $\gamma$ as a policy setting and do not assert that it is the largest feasible value for every scenario.

}

\subsection{Minimizing Total Travel Distance} \label{sec:min-total-distance}

\rev{We now develop an integer programming model (TTD) that groups teams into divisions and conferences such that the total travel distance across all teams is minimized. The model is based on the road trip distance measure introduced in Section~\ref{sec:measure-travel-distance}. It uses decision variables $x_{ij}$ which are 1 if teams $i$ and $j$ are in the same division, 0 otherwise, and $y_{ij}$ which equal 1 if teams $i$ and $j$ are in the same conference, 0 otherwise. We use symmetric pair variables, writing $x_{ji}=x_{ij}$ and $y_{ji}=y_{ij}$ for $i<j$. The variables $z_{ip}$ reflect the number of times that trip $p$ is used by team $i$.
Further, let $s_D$ be the number of teams in a each division (currently, $s_D=5$) and $s_C=|T|/2$ the number in a conference (currently, $s_C=15$); we assume equal-sized divisions and equal-sized conferences in all settings in this paper. 

\begingroup
\allowdisplaybreaks[4]
\begin{align}
\text{[TTD]} \quad \text{min}  \quad &\sum_{i \in T} r_i&&\label{eq:obj-mintravel}\\
\text{s.t.} \quad &r_i = \frac{1}{H}\sum_{p\in P_i}c_{ip}z_{ip} &&\forall i\in T,\label{eq:distance-def1}\\
&\sum_{j\ne i}x_{ij}=s_D-1
&&\forall i\in T,\label{eq:division-constraint}\\
&\sum_{j\ne i}y_{ij}=s_C-1
&&\forall i\in T,\label{eq:conference-constraint}\\
&x_{ij}+x_{jk}\le1+x_{ik} &&\forall  i,j,k \in T,  \label{eq:transitivity-x-one}\\
&y_{ij}+y_{jk}\le1+y_{ik} &&\forall  i,j,k  \in T, \label{eq:transitivity-y-one}\\
&x_{ij}\le y_{ij}
&&\forall i<j \in T,\label{eq:division-conference-relation}\\
&\sum_{p\in P_i}\delta_{jp}z_{ip}=a_1x_{ij}+a_2(y_{ij}-x_{ij})+a_3(1-y_{ij})
&&\forall i\ne j \in T,\label{eq:trip-cover}\\
&H\,B^{\min}\le\sum_{p\in P_i}z_{ip}\le H\,B^{\max}
&&\forall i\in T,\label{eq:trip-count-bounds}\\
& \sum_{j \in T\setminus\{i\}} m_j \left(a_1 x_{ij} + a_2 (y_{ij} - x_{ij}) + a_3 (1 - y_{ij}) \right) \geq \gamma \cdot \Bar{M}_i && \forall i \in T^I \label{eq:market-constraint} \\
&x_{ij},y_{ij}\in\{0,1\}
&&\forall i<j \in T,\label{eq:binary-vars}\\
&z_{ip}\in\mathbb Z_{\ge0}
&&\forall i\in T,\ p\in P_i.
\label{eq:trip-integer-vars}
\end{align}
\endgroup

The objective function (\ref{eq:obj-mintravel}) minimizes the total road trip distance across all teams, as defined in Constraints~(\ref{eq:distance-def1}). The required number of teams per division and conference is settled in Constraints~(\ref{eq:division-constraint}) and (\ref{eq:conference-constraint}), respectively. 
}
Constraints \eqref{eq:transitivity-x-one}–\eqref{eq:division-conference-relation} enforce logical grouping relationships within and among divisions and conferences.
\rev{
Constraints~\eqref{eq:trip-cover} link the required number of visits to each opponent to the selected road trips. Constraints~\eqref{eq:trip-count-bounds} bound each team's total number of road trips over the $H$-season horizon between $H\,B^{\min}$ and $H\,B^{\max}$. We set $(B^{\min},B^{\max})=(17,24)$ for 82-game seasons to limit both excessive consolidation into long tours and excessive fragmentation into short trips. For 72-game seasons, scaling these bounds by $72/82$ and rounding to the nearest integer gives $(15,21)$. These bounds restrict the average annual number of road trips over the horizon, rather than the number in each individual season.}
Finally, the Constraints~\eqref{eq:market-constraint} guarantee that each team retains at least $\gamma$ percent of its original market exposure, preserving revenue opportunities after expansion. 
\rev{The binary and integer domains are enforced by Constraints~\ref{eq:binary-vars} and \ref{eq:trip-integer-vars}, respectively.


}

\rev{
\subsection{The Kalai--Smorodinsky (KS) proportional-improvement model}
\label{sec:ks-model}

%
Minimizing total travel does not directly control how travel is distributed across teams. We therefore introduce a second model that explicitly considers each team’s travel burden and seeks to reduce it as much as possible. Following the proportional-improvement principle of \citet{kalai1975other}, the KS model seeks an alignment that gives every team as large a fraction as possible of its potential travel reduction. Defining this fraction requires two benchmarks for each team: an ideal distance and a reference distance.
}

We define an \textit{ideal travel distance} ($\hat{r}_i$) for each team $i$, which is the minimum possible travel distance team $i$ would face under a perfectly localized assignment, ignoring the preferences of all other teams. In other words, each team independently selects its in-division, in-conference (but out-of-division), and out-of-conference opponents solely to minimize its own total travel distance, without regard to league-wide consistency or mutual compatibility. Values for team $i$'s ideal travel distance can be obtained by solving a relaxation of the TTD model with an objective that minimizes only team $i$'s travel, \rev{while ignoring Constraints  (\ref{eq:distance-def1}), (\ref{eq:trip-cover}), (\ref{eq:trip-count-bounds}), and (\ref{eq:market-constraint}) for all other teams than $i$.}


\rev{
We also need to specify a reference distance $B_i>\hat r_i$, representing an acceptable, and potentially far larger than ideal, travel benchmark for team $i$. Incumbent references are based on travel under the current alignment, scaled by $G/82$ for a $G$-game season, allowing a pre-specified tolerance for computational purposes. Expansion teams have no current travel benchmark; their references are based on the largest incumbent travel-to-ideal ratio. Reference and ideal distances are fixed before optimizing each scenario.

For a proposed alignment with team travel $r_i$, define the normalized improvement
\begin{equation}
\phi_i=\frac{B_i-r_i}{B_i-\hat r_i}.
\label{eq:ks-gain}
\end{equation}
The numerator is the reduction from the reference distance, and the denominator is the gap between ideal and reference distances. Thus, $\phi_i=0.6$ means that team $i$ achieves 60\% of this potential reduction, rather than reducing its travel by 60\%.

Let $\lambda$ be a lower bound on every team's normalized improvement. The KS model is
\begin{align}
\text{[KS]}\qquad
\max\quad &\lambda,
\label{eq:obj-ks}\\
\text{s.t.}\quad
&\text{Constraints~(\ref{eq:distance-def1})--%
(\ref{eq:trip-integer-vars})},
\nonumber\\
&\lambda \le \frac{B_i-r_i}{B_i-\hat r_i},
&&\forall i\in T,
\label{eq:ks-fraction}\\
&0\le\lambda\le1.
\label{eq:ks-domain}
\end{align}
Constraint \eqref{eq:ks-fraction} bounds $\lambda$ from above by team $i$'s normalized improvement $\phi_i$. Maximizing $\lambda$ therefore maximizes the smallest normalized improvement across all teams. The constraints are linear because $B_i$ and $\hat r_i$ are fixed constants.

KS balances proportional improvements, rather than requiring equal mileage or guaranteeing that every team improves on its current travel. We use the KS principle with prescribed references and separately computed ideals, without claiming the full classical bargaining characterization.
}

\section{Results}
\label{sec:results}

In this section, we evaluate the current NBA structure and propose a flexible realignment framework with two objectives: minimizing total travel distance and ensuring fairness through the KS model.
%
\rev{We first compare the current 30-team alignment using both models. We then consider 32-team expansion scenarios using KS as the principal fairness model.
Throughout our numerical experiments, we use a time limit of 3,600 seconds for our instances.
We retain feasible solutions when optimality is not established within the allocated budget.}

\subsection{Current NBA Structure}
\label{sec:current-structure}

While the current structure has remained largely consistent over the past two decades, it may not be optimal when evaluated through the lens of travel efficiency or market equity.
%
\rev{We use each team's individual ideal travel as a benchmark. This is obtained from a team-specific relaxation within the retained trip pool; the resulting individual choices need not form a feasible league alignment. It allows us to compare relative travel burdens while accounting for differences in team locations.}

\begin{figure}[!htbp]
\centering
\subfloat[Current NBA divisional structure (30 teams)]{
\includegraphics[width=0.6\textwidth]{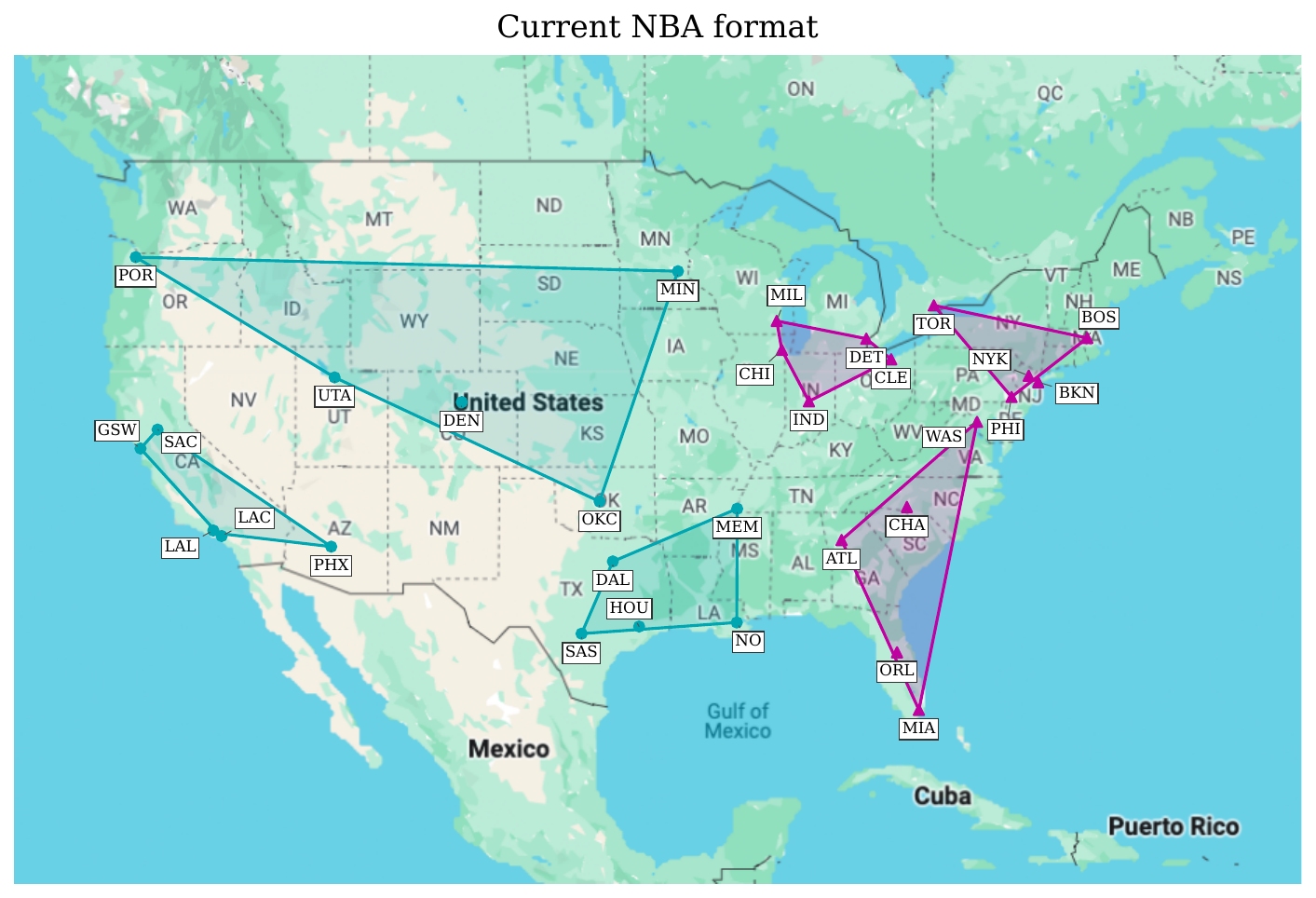}\label{fig:current-structure-visuals-a}
}
\hfill
\subfloat[Increase in travel distance vs.\ ideal for each team]{
\includegraphics[width=1\textwidth]{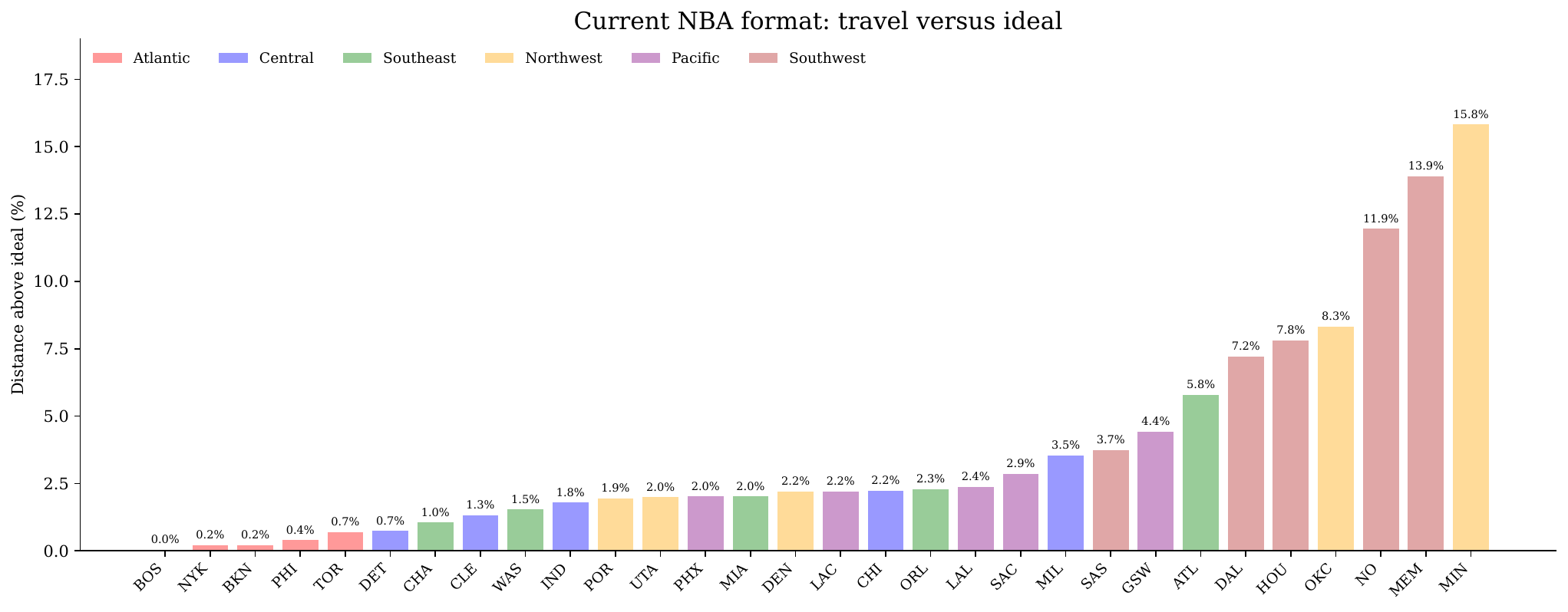}\label{fig:current-structure-visuals-b}
}
\caption{Current NBA structure: (a) divisional layout; (b) team travel inefficiencies relative to ideal travel.}
\label{fig:current-structure-visuals}
\end{figure}

%
Figure~\ref{fig:current-structure-visuals} illustrates the current NBA format from two perspectives. Figure \ref{fig:current-structure-visuals-a} maps the geographical distribution of teams within their existing divisional boundaries. \rev{Figure \ref{fig:current-structure-visuals-b} shows each team's percentage travel above its individual ideal, calculated as $100(r_i^0/\hat r_i^0-1)$. The bars and their labels report this percentage. Boston lies close to its ideal, whereas Minnesota, Memphis, and New Orleans are approximately 15.8\%, 13.9\%, and 11.9\% above theirs, respectively. 
}

The wide variation across franchises highlights the uneven distribution of travel demands, reinforcing the need to explore realignment alternatives. Together, these visualizations establish the empirical foundation for evaluating our optimization models, which aim to reduce inefficiencies and improve fairness by rethinking the league’s structural design.

\subsection{Evaluating Optimization Models on the Current NBA Structure}
\label{sec:current-30team-optimization}


\rev{To evaluate the realignment potential of our models, we apply the minimum-total-travel model described in Section~\ref{sec:min-total-distance} and the KS model in Section~\ref{sec:ks-model} to the current 30-team format. This provides a comparison with the league's existing structure in Figure~\ref{fig:current-structure-visuals-a}.}

\begin{figure}[!htbp]
\centering
\subfloat[Total Distance Minimization]{
\includegraphics[width=0.48\textwidth]{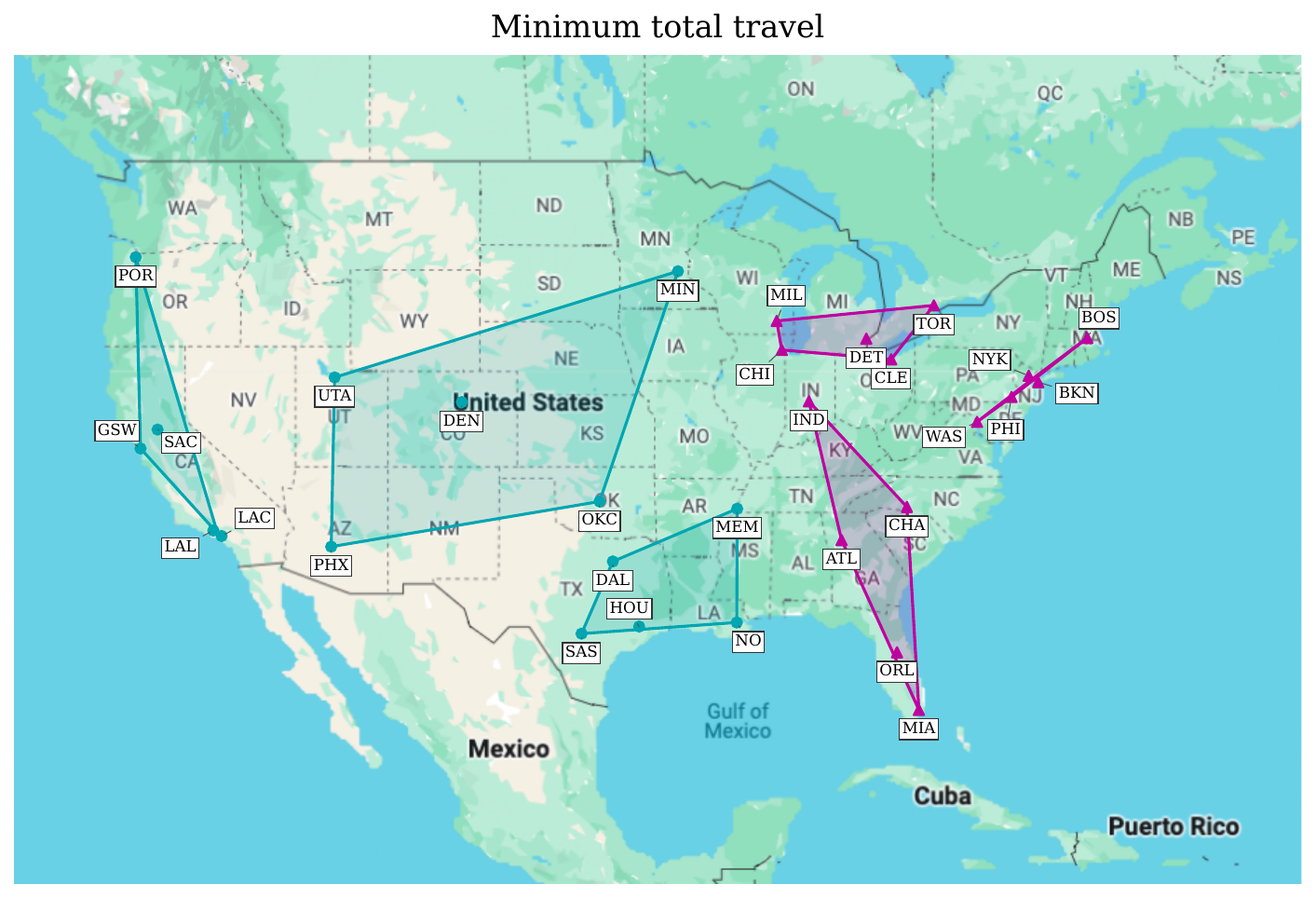}
}
\hfill
\subfloat[KS Model]{
\includegraphics[width=0.48\textwidth]{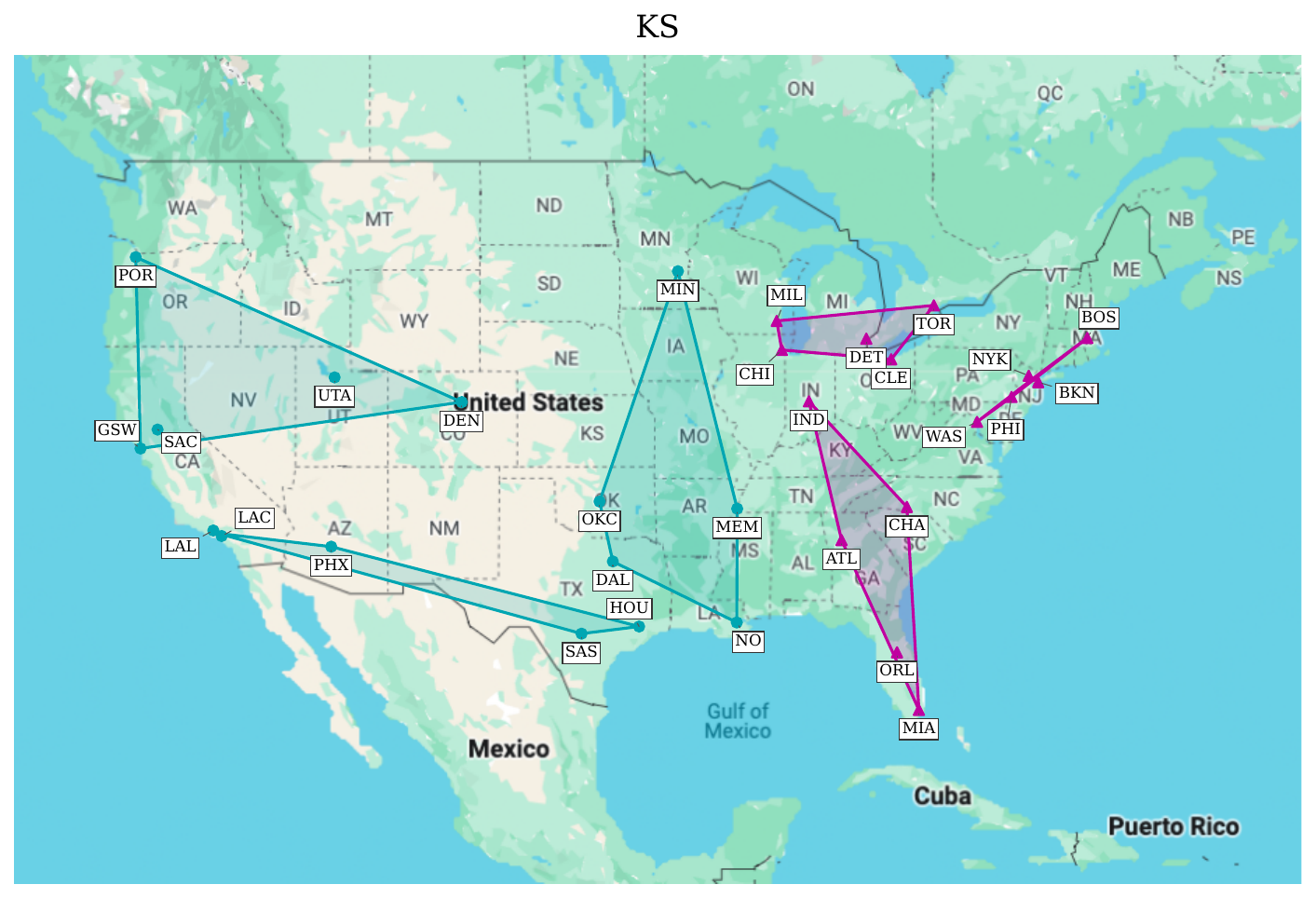}
}
\caption{Realignment of NBA divisions using two different optimization models applied to the existing 30-team structure.}
\label{fig:optimized-30team-layouts}
\end{figure}

Figure~\ref{fig:optimized-30team-layouts} presents the outcomes of these two optimization strategies. The left panel shows the divisional structure from minimizing total travel distance, while \rev{the right panel shows the KS alignment, which maximizes the smallest normalized reference-to-ideal travel gain across teams.
}

%
%
\rev{Both objectives produce the same Eastern divisions but different Western divisions. The maps illustrate how changing the objective can alter the alignment, while the travel and normalized gain measures below quantify the resulting efficiency-fairness tradeoff.}
Another important observation is that certain teams, because of their isolated locations, will always pose challenges for division-based designs. The addition of expansion teams may ease these challenges by offering more flexibility in balancing divisions.

%
\rev{The compact northeastern division appears under both objectives. The distinction between the models therefore concerns their treatment of team travel burdens, rather than a guarantee that KS produces divisions of similar geographic size. 
}

%
\rev{In terms of aggregate outcomes, the minimum-total-travel solution yields approximately 957{,}580 miles per year, compared with 958{,}513 miles under the current alignment. The fact that the current alignment is so close to the optimum suggests that minimizing travel is a primary objective for the NBA when structuring its conferences and divisions. The saved KS solution yields 960{,}514 miles, approximately 0.31\% above the minimum-total-travel solution. Its smallest normalized gain is 0.537, compared with 0.525 for the minimum-total-travel solution. The worst travel-to-ideal ratio falls from 1.158 to 1.154. Thus, the saved KS solution improves the normalized fairness measure at a modest increase in total travel. 
}

%
\rev{These results illustrate the trade-off between aggregate efficiency and protection of individual normalized gains. Accordingly, we use KS as the principal fairness model in the expansion analysis below.}

\rev{\subsection{Expansion Scenarios: Applying the KS Model}
\label{sec:ks-expansion-results}}

%
%
\rev{We examine 15 expansion scenarios, pairing two teams among six candidate cities: Seattle, Las Vegas, Vancouver, Montreal, Mexico City, and Tampa. Each pair is evaluated under four configurations, summarized in Table~\ref{tab:experiment-configurations}. Our primary configuration has 82 games per team and four divisions per conference. The alternatives change either the divisional format, the season length, or both.
In these numerical experiments, the median minimum normalized gain across all 60 configurations was 0.627, and 52 configurations (86.7\%) achieved a KS objective value of at least 0.50, meaning that every team attained at least half its reference-to-ideal travel reduction. Optimality was not established for all configurations and the reported outcomes in this section represent achieved performance under 3,600 CPU time limit, conditional on the retained candidate-trip pools. Relative optimality gap range between 8\% to 20\% for majority of the instances.
}

\begin{table}[!htbp]
\centering
\footnotesize
\caption{Expansion configurations included in the analysis.}
\label{tab:experiment-configurations}
\begin{tabular}{rrrr}
\toprule
Games per team &
\shortstack{Divisions per conference} &
Teams per division &
\shortstack{Expansion scenarios} \\
\midrule
82 & 4 & 4 & 15 \\
82 & 2 & 8 & 15 \\
72 & 4 & 4 & 15 \\
72 & 2 & 8 & 15\\
\bottomrule
\end{tabular}
\end{table}
\rev{The rotation horizons are $H=12$ for the 82-game format with four divisions per conference, $H=8$ for the 82-game format with two divisions, $H=24$ for the 72-game format with four divisions, and $H=16$ for the 72-game format with two divisions.}

%
\begin{figure}[!htbp]
    \centering
\includegraphics[width=\textwidth]{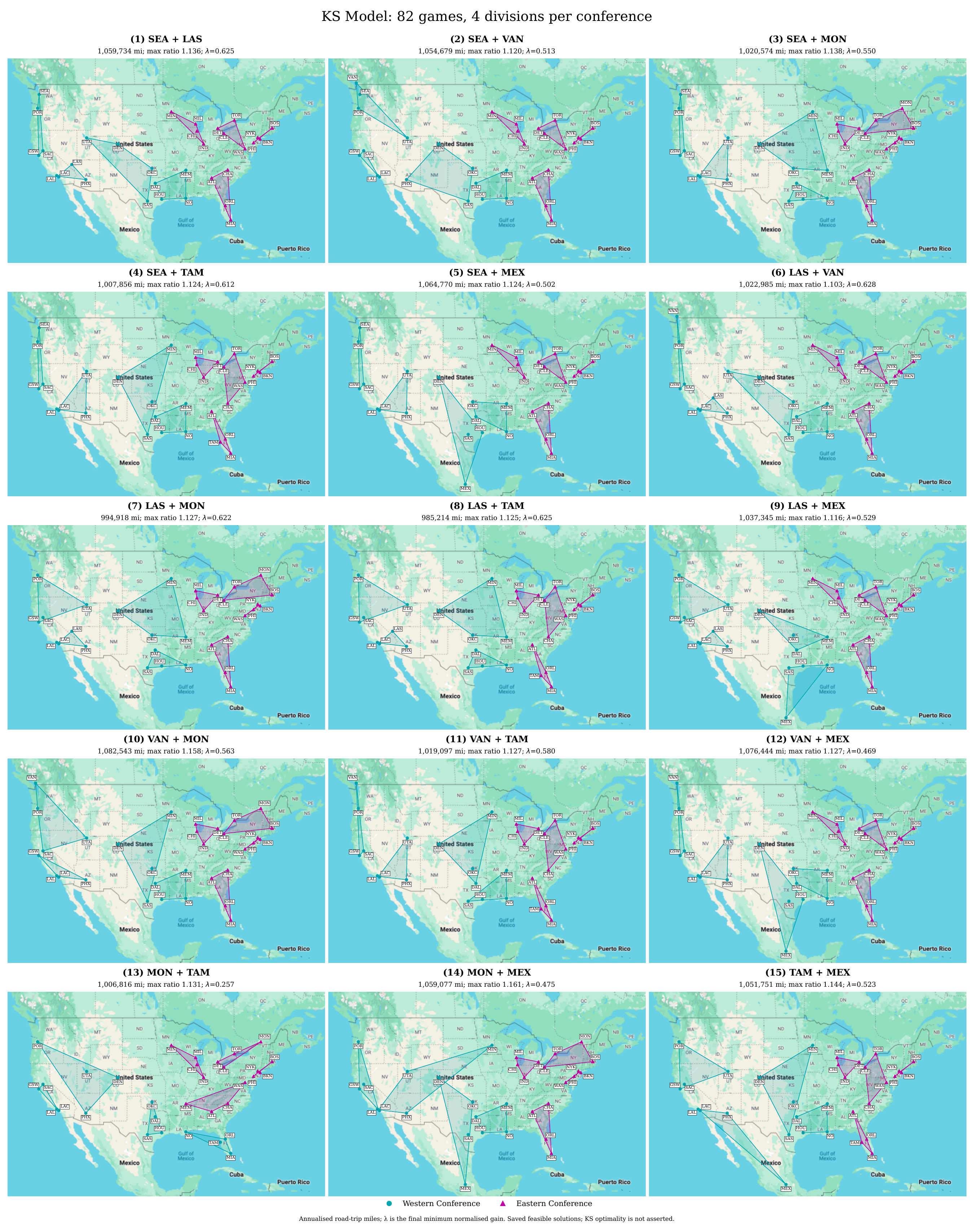}
\caption{KS divisional structures for all 15 expansion scenarios
(82-game season, four divisions per conference). \rev{Entries report annualized road-trip miles, the largest team travel-to-ideal ratio, and the reported common normalized gain level $\lambda$.}}
    \label{fig:KS_82tg_4divNo}
\end{figure}

%
\rev{Figure~\ref{fig:KS_82tg_4divNo} presents the KS alignments for all 15 city pairs under the primary configuration. It also reports total annualized road-trip travel across all 32 teams. Among the saved solutions, totals range from 985{,}214 miles for Las Vegas--Tampa to 1{,}082{,}543 miles for Vancouver--Montreal. Vancouver--Mexico City also has a relatively high total, at 1{,}075{,}503 miles. These results show that the effect of expansion depends on the city pair and the resulting alignment. A candidate's location alone does not determine league-wide travel.}

%
\rev{The team-level metrics complement these totals. The largest travel-to-ideal ratio identifies the most burdened team relative to its individual ideal, while $\lambda$ reports the common reference-to-ideal gain level retained by the solve. For example, a ratio of 1.136 means that the highest relative travel burden is 13.6\% above the corresponding ideal. A gain level of 0.625 represents 62.5\% of the reference-to-ideal improvement, not a 62.5\% reduction from current travel. The maps illustrate the alignments associated with these trade-offs; geographic compactness is not itself guaranteed by the KS objective.}
These insights illustrate how expansion decisions carry both logistical and strategic implications, underscoring the value of data-driven analysis in shaping the NBA’s future structure.

\rev{
\subsubsection{Seasonal realization}

%
To examine whether the aggregate travel estimates remain representative when visits are allocated to individual seasons, we fix one KS alignment for each of the 15 expansion pairs under the 82-game, four divisions per conference format. For each alignment, we generate ten distinct feasible twelve-season allocations, giving 150 rotations in total. A separate integer model uses randomized objective coefficients to vary the allocation of opponent matchups and home/away visits across seasons, while preserving the aggregate visit requirements and 41 home and 41 away games per team per season. We then optimize road-trip groupings for these seasonal visit requirements. Thus, the variation arises from alternative seasonal allocations, not from repeated optimization of the KS alignment.

Figure~\ref{fig:ks-seasonal-variation} shows that seasonal travel varies modestly around each team's realized rotation mean, with a mean absolute deviation of 1.32\%.
Two separate comparisons, not plotted in the figure, use the aggregate model estimates as their baseline: realized annual league travel is 1.98\% higher on average, and the mean absolute team-season percentage deviation is 2.95\%. These modest average discrepancies support the aggregate formulation for strategic realignment analysis in the tested cases, although individual deviations can be larger. The construction does not establish feasibility with respect to calendar dates, arena availability, or rest requirements.
\begin{figure}[!htbp]
    \centering
    \includegraphics[width=1\textwidth]
        {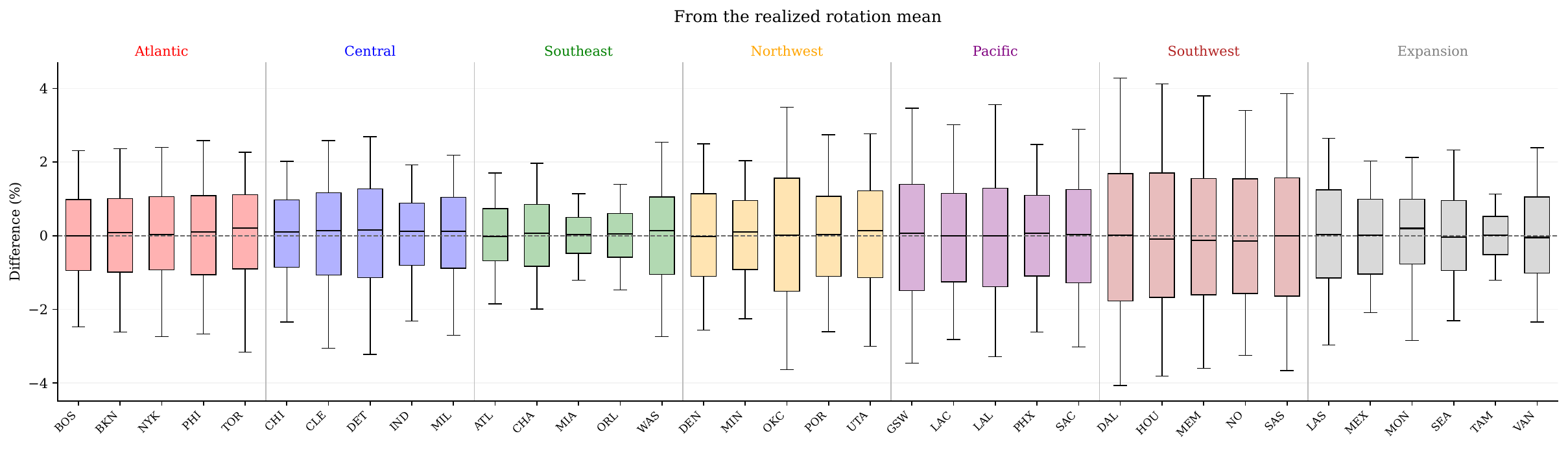}
    \caption{Seasonal travel variation for the KS alignments under the 82-game format with four divisions per conference. Values are percentage deviations from each team's mean travel within its realized twelve-season rotation, pooled across the tested scenarios and rotations. Boxes show the interquartile range, horizontal lines the medians, and whiskers the 5th and 95th percentiles. Observations beyond the whiskers are omitted.}
    \label{fig:ks-seasonal-variation}
\end{figure}
}

\subsection{Alternative Competition Design Scenarios}
\label{sec:alternative-competition-design}

While our primary analysis assumes the traditional 82-game season and four divisions per conference, recent debates in the NBA have raised alternative competition designs worth exploring. 

The 82-game season, unchanged since 1967–68, faces criticism for player fatigue and load management \citep{lewis2018hardknock}. Many players and coaches support reducing the schedule to 72 games, but the NBA fears the revenue loss from fewer games. Expansion, however, offers a compromise as 32 teams $\times$ 72 games yields 1,152 games, close to today’s 1,230.

%
%
\rev{We compare the primary configuration with the two alternatives listed in Table~\ref{tab:experiment-configurations}: two divisions per conference with 82 games, and four divisions per conference with 72 games.}

\begin{figure}[!htbp]
    \centering
\includegraphics[width=\textwidth]{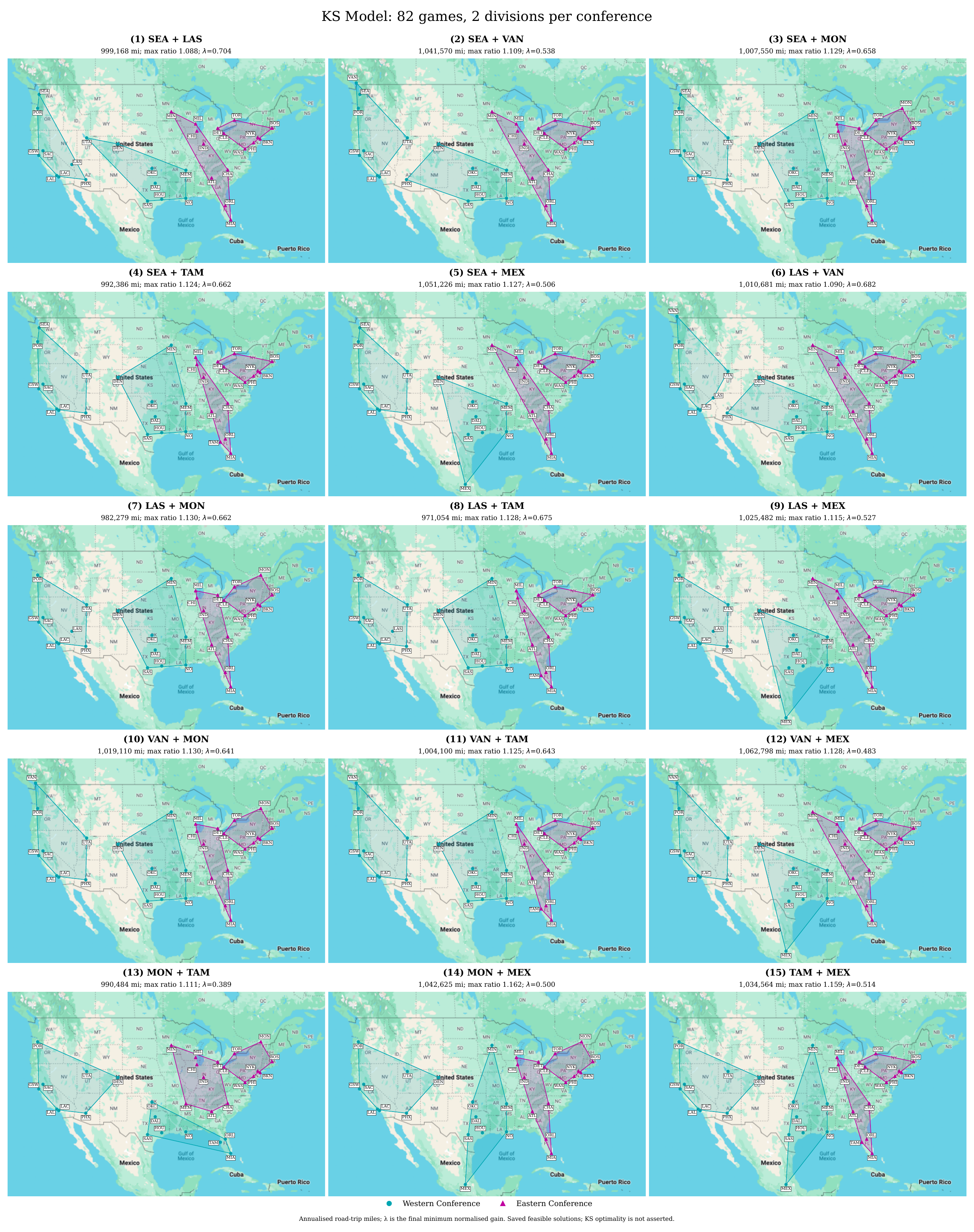}
    \caption{KS divisional structures for all 15 expansion scenarios (82-game season, two divisions per conference). \rev{Entries report annualized road-trip miles, the largest team travel-to-ideal ratio, and the reported common normalized gain level $\lambda$.}}
    \label{fig:KS_82tg_2dvNo}
\end{figure}

\begin{figure}[!htbp]
    \centering
\includegraphics[width=\textwidth]{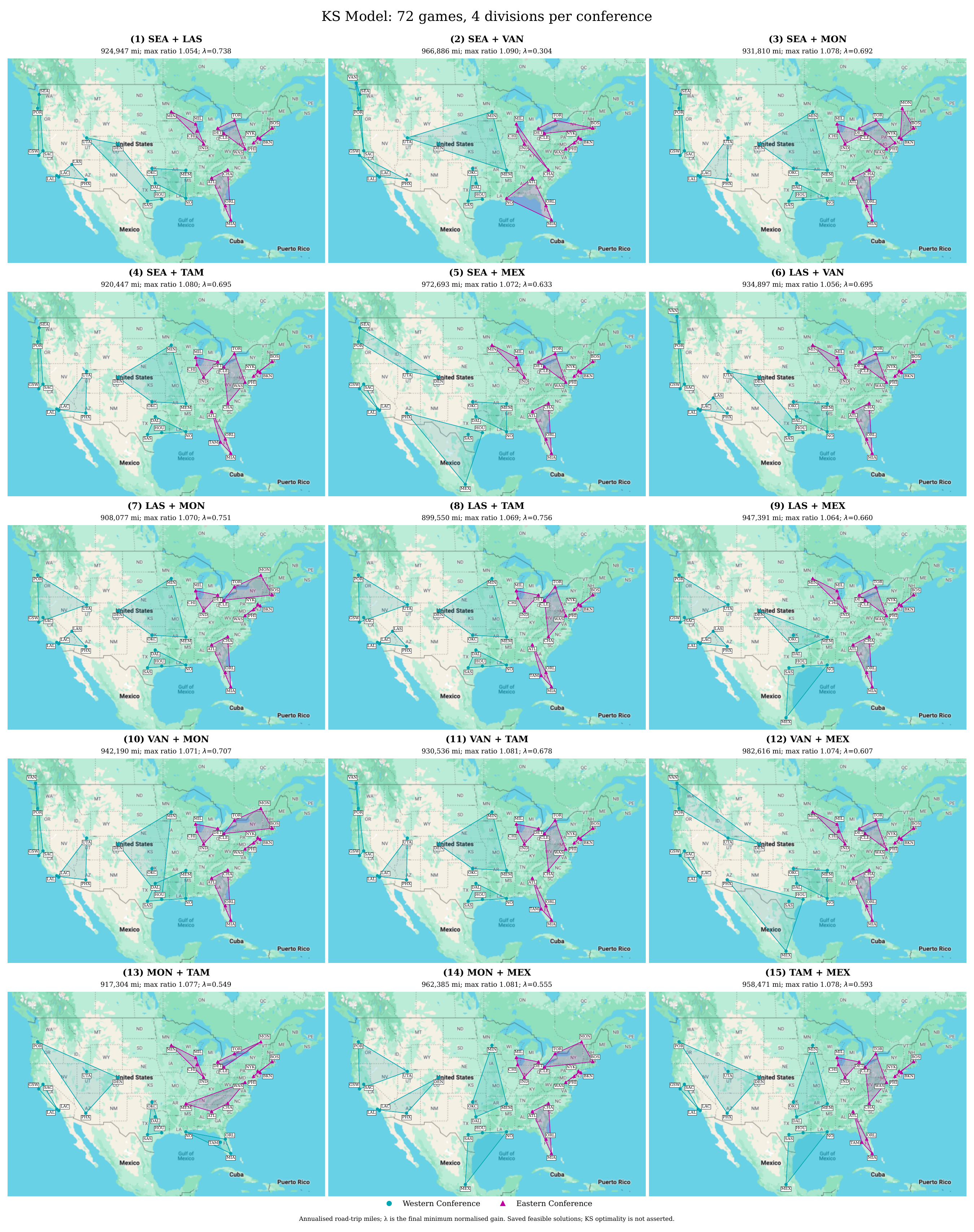}
    \caption{KS divisional structures for all 15 expansion scenarios (72-game season, four divisions per conference). \rev{Entries report annualized road-trip miles, the largest team travel-to-ideal ratio, and the reported common normalized gain level $\lambda$.}}
    \label{fig:KS_72tg_4dvNo}
\end{figure}

%
\rev{Figures~\ref{fig:KS_82tg_2dvNo} and~\ref{fig:KS_72tg_4dvNo} present the alternative formats. In the saved 82-game solutions, two divisions per conference yield lower total travel than four in 14 of the 15 matched city pairs; Montreal--Tampa is the exception. The 72-game, four-division solutions have lower totals than their corresponding 82-game solutions, reflecting both fewer games and the selected alignments. These comparisons concern the obtained solutions and do not establish that one format is globally preferable.}

%

%

\begin{figure}[!htbp]
    \centering
\includegraphics[width=\textwidth]{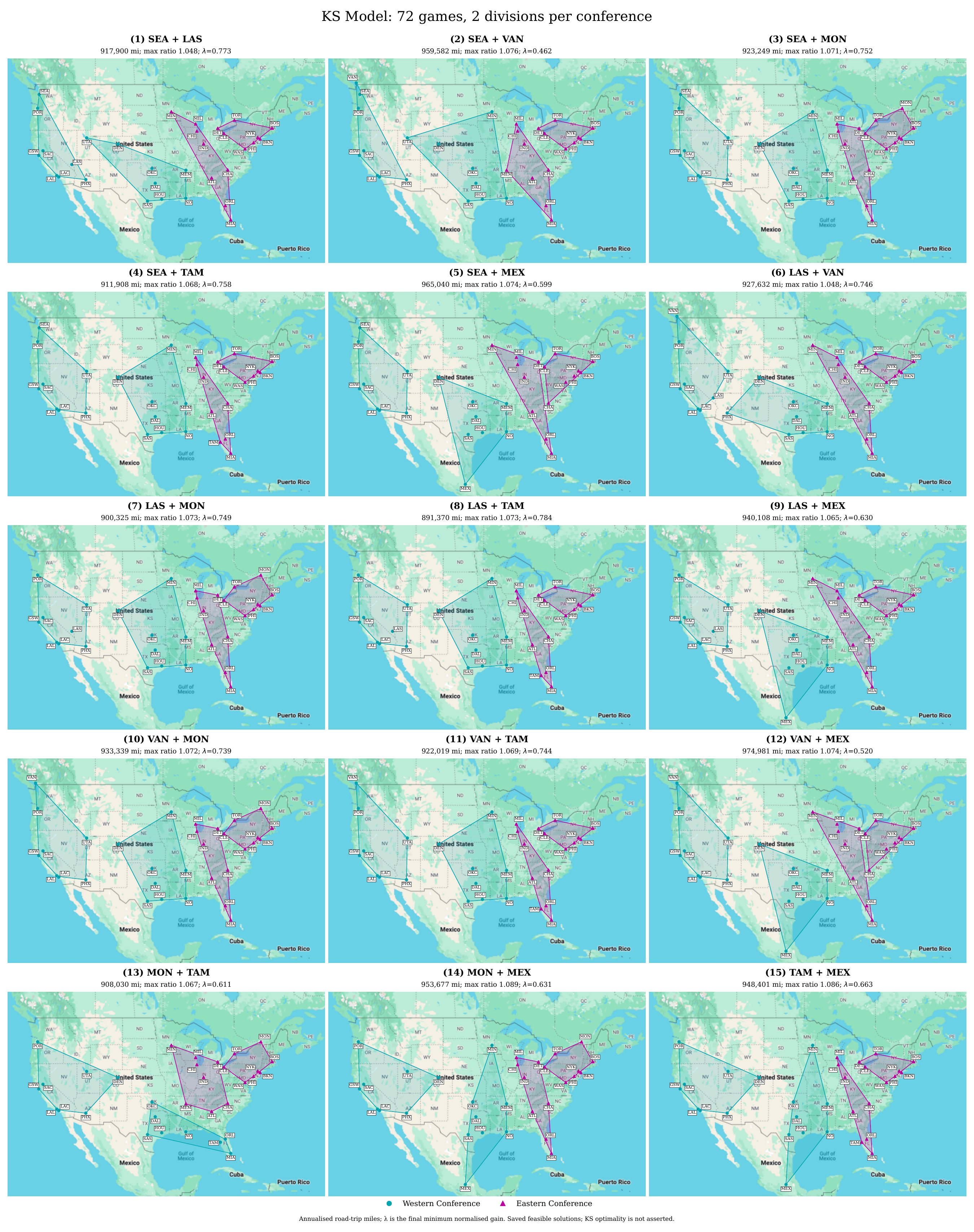}
    \caption{KS divisional structures for all 15 expansion scenarios (72-game season, two divisions per conference). \rev{Entries report annualized road-trip miles, the largest team travel-to-ideal ratio, and the reported common normalized gain level $\lambda$.}}
    \label{fig:KS_72tg_2dvNo}
\end{figure}

%

\section{Discussion and Concluding Remarks}
\label{sec:discussion-conclusion}

In this study, we examined how the NBA could restructure its conferences and divisions in light of a possible expansion to 32 teams. By applying optimization techniques, we evaluated alternative structures that balance travel efficiency, fairness, and \rev{opponent exposure}. Beyond producing solutions, the models provide a flexible decision support tool that allows league executives to test different expansion scenarios and assess their consequences.

Our findings suggest several practical takeaways for NBA executives and policymakers.
%
\rev{First, minimizing total annualized road-trip travel provides an efficiency benchmark for comparing realignments. The results show that season length, division structure, and expansion locations should be assessed jointly; no single format is uniformly preferable across all criteria.}
%
\rev{Second, the KS model offers a principled way to protect the team with the smallest normalized travel improvement between its reference and individual ideal. Reporting this gain alongside total travel makes the efficiency--fairness trade-off explicit.}
%
\rev{Finally, the choice of expansion cities must consider both commercial considerations and geographic fit. The travel consequences depend on the city pair and the resulting league alignment, so candidates should be evaluated jointly rather than solely on their individual locations.}
%
\rev{The market-exposure constraint requires each incumbent to retain at least 80\% of its baseline valuation-weighted opponent exposure. This provides an explicit exposure safeguard, but does not guarantee a corresponding level of revenue.}

Several limitations provide avenues for future work.
%
\rev{Our analysis focuses on the long-term design of conferences and divisions, representing opponent visits and road trips over a multi-season rotation. The seasonal realization experiment examines whether the selected alignments can support integer annual game and visit allocations, with modest average travel discrepancies in the tested cases. While this seems are reasonable approximation, we do not construct a dated league schedule. Future work could therefore integrate arena availability, rest requirements, back-to-back games, and synchronization of the teams' trips.}
In addition, travel distance alone does not fully capture fatigue or performance impacts, which could be better addressed by integrating sports science data.
%
\rev{Franchise valuations provide weights for the opponent-exposure measure, but the model does not estimate actual revenue. Future work could incorporate broadcast ratings, attendance forecasts, and sponsorship data to evaluate the financial consequences of realignment more directly.}
%
\rev{
The computational results are also conditional on the set of road-trip candidates, and some of the obtained solutions lack certified optimality. Larger candidate pools and further search may improve the reported solutions.}

%
\rev{NBA expansion decisions must balance commercial opportunity, operational feasibility, and fairness across teams.}
Our analysis shows that optimization-based tools can provide clear, data-driven guidance for navigating these trade-offs. We anticipate that such approaches will become increasingly important as professional leagues grow more global, competitive, and commercially complex.
%
\rev{The framework can also be adapted to other professional leagues facing expansion or realignment, provided that their competition rules, travel patterns, and exposure measures are represented appropriately.}


{\footnotesize
\bibliographystyle{elsarticle-num-names}
\bibliography{references}
}

\end{document}